\documentclass[a4,11pt]{article}
\usepackage{amsmath,amssymb}
\newtheorem{theorem}{Theorem}
\newtheorem{collorary}{Collorary}

\def\proof{\noindent {\bf Proof.}\quad}

\def\Qed{\hfill$\square$}
\def\rank{\mbox{\rm rank}}
\def\fool{\mbox{\rm fool}}

\def\bp{\mbox{\rm bp}}
\def\rank{\mbox{\rm rank}}
\def\fool{\mbox{\rm fool}}
\begin{document}
\title{Ordered Biclique Partitions and Communication Complexity Problems}
\author{Manami Shigeta\footnote{Dept. of Comput. Sci., Gunma Univ, 
Tenjin 1-5-1, Kiryu, Gunma 376-8515, Japan}
\qquad Kazuyuki Amano\footnotemark[1] \footnote{Corresponding author:
Email: {\tt amano@gunma-u.ac.jp}}}
\maketitle

\begin{abstract}
An {\it ordered biclique partition} of the complete graph $K_n$
on $n$ vertices is a collection of bicliques (i.e., complete bipartite graphs)
such that (i) every edge of $K_n$ is covered by at least one and
at most two bicliques in the collection, and 
(ii) if an edge $e$ is covered by two bicliques then each endpoint 
of $e$
is in the first class in one of these bicliques and in the second class
in other one.
In this note,
we give an explicit construction of such a collection of
size $n^{1/2+o(1)}$,
which improves the $O(n^{2/3})$ bound shown in the previous
work \cite[Disc. Appl. Math., 2014]{Ama13}.

As the immediate consequences of this result, we show
(i) a construction of $n \times n$ 0/1 matrices of
rank $n^{1/2+o(1)}$ which have a fooling set
of size $n$, i.e., the gap between rank and fooling set size 
can be at least almost quadratic, and
(ii) an improved lower bound $(2-o(1)) \log N$ on 
the nondeterministic communication complexity 
of the clique vs. independent set problem, which matches 
the best known lower bound on the deterministic
version of the problem shown by Kushilevitz, Linial and Ostrovsky
\cite[Combinatorica, 1999]{KLO99}.

{\bf keywords} biclique partition, Boolean matrix, fooling set, rank, complete graphs
\end{abstract}

\section{Introduction}
Let $G = (V,E)$ be an undirected graph.
For two disjoint subsets $U$ and $W$ of $V$,
the complete bipartite graph with edge set $U \times W$ is
called a {\it biclique} and is denoted by ${\mathcal B}(U,W)$.
For an integer $k \geq 1$,
a collection of bicliques $\{{\mathcal B}(U_i,W_i)\}_i$
is called a {\it $k$-biclique covering} of $G$ if every
edge in $G$ lies in at least one and at most $k$ bicliques in the 
collection.
The minimum size of a $k$-biclique covering of $G$ is denoted by
$\bp_k(G)$. 
In particular, a $1$-biclique covering is called a
{\it biclique partition} and its minimum size $\bp_1(G)$ is 
just denoted by $\bp(G)$.
Biclique coverings of graphs have been widely investigated
in the literature (see e.g., \cite{Alo96,GP72,JK09}).

In the preceding work \cite{Ama13}, we introduced
an ``intermediate" notion between the biclique partition and $2$-biclique
covering,
which we call an {\it ordered biclique partition}.
An ordered biclique partition of $G$ is a $2$-biclique
covering $\{{\mathcal B}(U_i,W_i)\}_i$ with an additional restriction 
that if an edge $e=\{u,v\}$ is
covered by two bicliques, say
${\mathcal B}(U_k,W_k)$ and ${\mathcal B}(U_\ell, W_\ell)$,
then each endpoint of $e$ belongs
to a distinct color class in these bicliques, i.e.,
$w \in U_k \cap W_{\ell}$ or $w \in U_{\ell} \cap W_k$ for $w \in \{u,v\}$.
The minimum size of such a partition is denoted by $\bp_{1.5}(G)$.
%Obviously from the definition, it holds that 
%$\bp_2(G) \leq \bp_{1.5}(G) \leq \bp(G)$.
Recently, in \cite{Ama13}, the second author of this note showed
$\bp_{1.5}(K_n) = O(n^{2/3})$ by giving an explicit construction
of such a partition, where $K_n$ denotes the complete graph
on $n$ vertices.

In this note, we improve this bound to $\bp_{1.5}(K_n) = n^{1/2+o(1)}$,
which is the main contribution of this note.
This bound is almost tight since $\bp_{1.5}(K_n) \geq \bp_{2}(K_n) = 
\Theta(n^{1/2})$ where the bound on $\bp_{2}(K_n)$ is due to 
Alon \cite{Alo96}.

The original motivation for considering such a parameter is its
close connection to the problems related to communication complexity.
One of such is the ``rank" vs. ``fooling set" problem.
Let $M$ be an $n \times n$ 0/1-matrix.
The rank of $M$ over the reals is denoted by $\rank(M)$.
A set $S \subseteq \{1,2,\ldots, n\} \times \{1,2, \ldots, n\}$ 
of the index set of $M$ is called
a {\it fooling set} for $M$ if there exists a value $z \in \{0,1\}$ such that
\begin{enumerate}
\item
for every $(k,\ell) \in S$, $M_{k,\ell}=z$,
\item
for any distinct $(k_1, \ell_1)$ and $(k_2, \ell_2)$ in $S$,
$M_{k_1,\ell_2}\neq z$ or $M_{k_2, \ell_1} \neq z$.
\end{enumerate}
The largest size of a fooling set of $M$ is denoted by $\fool(M)$.
Analyzing the size of a fooling set is one of the main tools for proving lower
bounds on the communication complexity (see e.g., \cite{KN97}).

It is known that $\fool(M) \leq (\rank(M)+1)^2$ (see Dietzfelbinger, Hromkovi\v{c} and Schnitger \cite{DHS96}).
The open question is whether this quadratic gap can be improved or not
(see e.g., \cite[Open Problem 2]{DHS96}).
M.~H\"{u}hne (described in \cite{DHS96}, and see also 
\cite{Ama13,The11}) constructed a matrix $M$ such that 
$\fool(M) \geq \rank(M)^{\log_4 6} = \rank(M)^{1.292\cdots}$.
This was improved to
$\fool(M) \geq \Omega(\rank(M)^{1.5})$ in the previous work
of the second author of this note \cite{Ama13}.
A biclique partition presented in this note immediately
gives a new separation $\fool(M) \geq \rank(M)^{2-o(1)}$, which is almost tight.
Note that recently Friesen and Theis \cite{FT12} proved that the exponent $2$
on the rank is tight if we take the rank in a field of 
characteristic two. 
See also \cite{HL13} for a recent development on a related problem.

Our new partition also
gives an improved bound
on the nondeterministic communication complexity 
of the clique vs.\ independent set problem (see e.g., a textbook
\cite{KN97} for the background and definition of the problem).
It was shown that
finding a graph $H$ with $\chi(H) \geq f(\bp_{1.5}(H))$ for some function
$f(\cdot)$ is essentially equivalent to proving $\log_2 f(N)$
lower bound on the nondeterministic communication complexity for 
the problem for an explicit graph on $N$ 
vertices, 
where $\chi(H)$ denotes the chromatic number of $H$ \cite{Ama13}.
(See also \cite{BLT13,Lag12} for this equivalence. In these papers,
$\bp_{1.5}(\cdot)$ is denoted by $\bp_{or}(\cdot)$.)
Combining this with our biclique partition
yields that the nondeterministic communication complexity
of the problem is at least $(2-o(1))\log_2 N$,
which improves the previously known bounds
of $1.5 \log_2 N$ in \cite{Ama13} and $1.2 \log_2 N$ in 
\cite{HS12} and matches the
best known lower bound on the {\it deterministic} version of 
the problem shown by Kushilevitz, Linial and Ostrovsky
shown in \cite{KLO99} (see also \cite{KW09}).

\section{Ordered Biclique Partition of Complete Graphs}

Let $[n]$ denote the set $\{1,2,\ldots, n\}$.
As defined in the Introduction, for an undirected graph $G$, 
$\bp_{1.5}(G)$ is the minimum size of an 
{\it ordered biclique partition} of $G$.
The following is an example of such a partition
$\{{\mathcal B}(U_i,W_i)\}_{i=1}^4$ of size four for $K_6$ on the
vertex set $[6]$:
\begin{eqnarray*}
U_1 = \{1,2\}, & & W_1 = \{4,6\},\\
U_2 = \{1,3\}, & & W_2 = \{2,5\}, \\
U_3 = \{3,6\}, & & W_3 = \{1,4\}, \\
U_4 = \{2,4,6\}, & & W_4 = \{3,5\}.
\end{eqnarray*}
The edges $\{1,6\}$, $\{2,3\}$ and $\{3,4\}$ are covered twice.
It can be checked that
$(1,6) \in (U_1 \times W_1) \cap (W_3 \times U_3)$,
$(2,3) \in (U_4 \times W_4) \cap (W_2 \times U_2)$ and
$(3,4) \in (U_3 \times W_3) \cap (W_4 \times U_4)$.
An easy case analysis verifies that $\bp_{1.5}(K_6)=4$.
%Note that $\bp(K_6)=5$ by the Graham-Pollak Theorem and
%$\bp_{2}(K_6)=3$ by the witness
%\begin{eqnarray*}
%U_1 = \{1,2\}, & & W_1 = \{3,4,5,6\},\\
%U_2 = \{3,4\}, & & W_2 = \{5,6\}, \\
%U_3 = \{1,3,5\}, & & W_3 = \{2,4,6\}.
%\end{eqnarray*}
%Hence these three parameters are all different for $K_6$.

In the previous work \cite{Ama13}, we showed
$\bp_{1.5}(K_n)=O(n^{2/3})$.
%In \cite{Ama13}, the second author of this note gave an explicit
%construction of ordered biclique partition of $K_{n}$ of
%size $O(n^{2/3})$.
The following theorem improves this result when
we put $k \geq 3$.

\begin{theorem}
\label{th:main}
${\rm bp}_{1.5}(K_{n^{2k-1}})=O(kn^k)$.
\end{theorem}

\proof
\quad
The theorem is obvious for $k=1$. %${\rm bp}_{1.5}(K_{n})=O(n)$.
For $k\geq 2$,
we consider the complete graph $K_{n^{2k-1}}$ 
on the vertex set $V=[n]^{2k-1}=\{(x_1, x_2, \ldots, x_{2k-1})\ |\ x_i\in[n]\}$.
Define three types of subsets of the edge set of $K_{n^{2k-1}}$:
\begin{eqnarray*}
	C_i&=&\{\{u,v\} \mid  u_{k+i}\neq v_{k+i} \mbox{ and } u_{i+\ell}=v_{i+\ell}\ (1\leq \ell\leq k-1)\} \mbox{ for } 0\leq i\leq k-1,\\
	D_j&=&\{\{u,v\} \mid u_{j}\neq v_{j} \mbox{ and } u_{k+j+\ell}=v_{k+j+\ell}\ (0\leq \ell\leq k-2)\} \mbox{ for } 1\leq j\leq k-1,\\
	E_{i,j}&=&\{\{u,v\} \mid  u_j\neq v_j \mbox{, } u_{k+i}\neq v_{k+i} \mbox{ and } u_{\ell}=v_{\ell}\ (1\leq \ell \leq j-1,\ k+j\leq \ell \leq k+i-1)\}\\
			&{}& \hspace{90mm}  \mbox{ for } 1\leq i \leq k-1,\ 1\leq j\leq i.
\end{eqnarray*}
Here the index ``$k+j+\ell$" % of the vertices $u_{k+j+\ell}$ and $v_{k+j+\ell}$ 
in the definition of $D_j$ is modulo $2k-1$.

For example, for $k=4$,
we define $C_i$'s for $i \in \{0,1,2,3\}$,
$D_j$'s for $j \in \{1,2,3\}$ and
$E_{i,j}$'s for $(i,j) \in \{(1,1),(2,1),(2,2),(3,1),(3,2),(3,3)\}$
as shown in Table \ref{tbl:k=4}.
In this table, 
`$\bigcirc$', `$\times$' and `$-$' denote $u_i=v_i$, $u_i\neq v_i$ and $don't\ care$, respectively.
%%%%%%%%%%%%%%%%%%
	%---------------------------------------------------------------------
	\begin{table}[htbp]
	\caption{$C_i$, $D_j$ and $E_{i,j}$ in the case $k=4$. For example,
$C_0$ is the set of edges such that the first three coordinates
(out of $2k-1=7$ in total) of its two endpoints are identical and the fourth
coordinates of them are different, which is represented by
``$\bigcirc \bigcirc \bigcirc \times - - -$''.}
  \label{tbl:k=4}
	\center
	$\begin{array}{|c|ccccccc|} \hline
		& x_1& x_2& x_3& x_4& x_5& x_6& x_7\\ \hline
  	C_0& \bigcirc& \bigcirc& \bigcirc& \times& -& -& -\\ \hline
	  C_1& -& \bigcirc& \bigcirc& \bigcirc& \times& -& -\\ \hline
	  C_2& -& -& \bigcirc& \bigcirc& \bigcirc& \times& -\\ \hline
		C_3& -& -& -& \bigcirc& \bigcirc& \bigcirc& \times\\ \hline \hline
	  D_1& \times& -& -& -& \bigcirc& \bigcirc& \bigcirc\\ \hline
	  D_2& \bigcirc& \times& -& -& -& \bigcirc& \bigcirc\\ \hline
	  D_3& \bigcirc& \bigcirc& \times& -& -& -& \bigcirc\\ \hline
	\end{array}$\hspace{5mm}
	$\begin{array}{|c|ccccccc|} \hline
		& x_1& x_2& x_3& x_4& x_5& x_6& x_7\\ \hline
	  E_{1,1}& \times& -& -& -& \times& -& -\\ \hline \hline
	  E_{2,1}& \times& -& -& -& \bigcirc& \times& -\\ \hline
	  E_{2,2}& \bigcirc& \times& -& -& -& \times& -\\\hline \hline
		E_{3,1}& \times& -& -& -& \bigcirc& \bigcirc& \times\\ \hline
	  E_{3,2}& \bigcirc& \times& -& -& -& \bigcirc& \times\\ \hline
	  E_{3,3}& \bigcirc& \bigcirc& \times& -& -& -& \times\\ \hline
	\end{array}$
	\end{table}
	%---------------------------------------------------------------------

We first see that the union of these subsets covers all 
edges in $K_{n^{2k-1}}$.
This can easily be verified by checking 
\begin{eqnarray*}
	\bigcup_i C_i &\supset& \{\{u,v\} \mid u \neq v \mbox{ and } u_{\ell}=v_{\ell}\ (1\leq \ell \leq k-1)\},
\end{eqnarray*}
and, for each $1\leq j\leq k-1$,
\begin{eqnarray*}
	D_j\cup \bigcup_i E_{i,j} &=& \{\{u,v\} \mid  u_{j}\neq v_{j} \mbox{ and } u_{\ell}=v_{\ell}\ (1\leq \ell \leq j-1)\}.
\end{eqnarray*}
%%%%%%%%%%%%%%%%%%

A key property of the collection of these subsets is that
among all pairwise intersections of subsets in the collection,
only $C_i\cap E_{i,j}$ is nonempty,
while the others are empty;
namely,
\begin{eqnarray}\label{eq:c cap e}
	C_i\cap E_{i,j}=\{\{u,v\} \mid u_j\neq v_j \mbox{, } u_{k+i}\neq v_{k+i}
								 \mbox{ and } u_{\ell}=v_{\ell}\ (1\leq \ell \leq j-1,\ i+1\leq \ell \leq k+i-1)\}
\end{eqnarray}
for $1\leq i\leq k-1,\ 1\leq j\leq i$.

%%%%%%%%%%%%%%%%%%
In order to construct an ordered biclique partition of $K_{n^{2k-1}}$,
we design a biclique partition of 
graph $G_{\mathcal{E}}=(V, \mathcal{E})$ 
for each $\mathcal{E}\in\{C_i, D_j, E_{i,j}\}_{i,j}$ separately.
We can observe that
$G_{C_i}$ (and also $G_{D_j}$)
is the $n^{k-1}$-blowup of
$n^{k-1}$ independent copies of $K_n$.
Here an $m$-blowup of a ``base" graph $H$ is obtained by replacing
every vertex of $H$ by a group of $m$ vertices
and every edge of $H$ by an $m\times m$ biclique between the corresponding groups of vertices.
We also observe that
$G_{E_{i,j}}$ is the $n^{2k-i-2}$-blowup of $n^{i-1}$ independent
copies of the complement of $n \times n$ grid graph $\overline{G}_{n,n}$.
Let $\widetilde{G}_{\mathcal{E}}$ denote the
base graph of $G_{\mathcal{E}}$, i.e.,
$\widetilde{G}_{C_i}$ and $\widetilde{G}_{D_j}$ 
are the $n^{k-1}$ independent copies of $K_n$ and
$\widetilde{G}_{E_{i,j}}$ is the
$n^{i-1}$ independent copies of $\overline{G}_{n,n}$.

%%%%%%%%%%%%%%%%%%
Two basic facts are needed to prove this theorem.
First, for any graph $H$ on the vertex set $\{v_1, \ldots, v_m\}$, ${\rm bp}(H)\leq m-1$.
This is because
the collection of $(m-1)$ stars $\{\mathcal{B}(\{v_i\}, N(v_i)\cap \{v_{i+1}, \ldots, v_m\})\}_{i=1}^{m-1}$
forms a biclique partition of $H$,
where $N(v_i)$ denotes the set of neighbors of $v_i$.
Second, if $H$ is a blowup of $\widetilde{H}$, then ${\rm bp}(H)\leq {\rm bp}(\widetilde{H})$.
The reason is that
the blowup of a biclique is a biclique itself;
the blowup of all the biclique in a partition of $\widetilde{H}$
is a biclique partition of $H$.
Because of these facts, we have
\begin{eqnarray*}
	{\rm bp}(G_{C_i}) &\leq& {\rm bp}(\widetilde{G}_{C_i})\leq n^{k-1}\cdot {\rm bp}(K_n)\leq n^{k-1}(n-1) \mbox{ for } 0\leq i\leq k-1,\\
  {\rm bp}(G_{D_j}) &\leq& {\rm bp}(\widetilde{G}_{D_j})\leq n^{k-1}\cdot {\rm bp}(K_n)\leq n^{k-1}(n-1) \mbox{ for } 1\leq j\leq k-1,\\
  {\rm bp}(G_{E_{i,j}}) &\leq& {\rm bp}(\widetilde{G}_{E_{i,j}})\leq n^{i-1}\cdot {\rm bp}(\overline{G}_{n,n})\leq n^{i-1}(n^2-1)
  					\mbox{ for } 1\leq i \leq k-1,\ 1\leq j\leq i.
\end{eqnarray*}
It would be worth noting that
we can slightly improve the upper bound on ${\rm bp}(\overline{G}_{n,n})$
although it affects only a lower order term.
If we place the vertices of $\overline{G}_{n,n}$ in an $n \times n$ 
square grid and
the roots of the stars are picked in row-major order
then the last row can be skipped.
Thus, $n(n-1)$ stars are enough to cover all edges
instead of a trivial bound of $n^2-1$,
i.e., ${\rm bp}(G_{E_{i,j}})\leq n^i(n-1)$.
Consequently, we obtain a collection of $(2k-1)\cdot n^{k-1}(n-1) + \sum_{i=1}^{k-1} i n^i(n-1)$ bicliques
that covers all edges in $K_{n^{2k-1}}$.

%%%%%%%%%%%%%%%%%%
To complete the proof,
we should notice that every edge $e \in C_i\cap E_{i,j}$ for $1\leq i\leq k-1,\ 1\leq j\leq i$
is covered by exactly {\it two} bicliques in the collection
(by recalling Eq.(\ref{eq:c cap e})).
% such as $\mathcal{B}(U_p, W_p)$ and $\mathcal{B}(U_q, W_q)$.
Therefore, in order to satisfy the definition of the ordered biclique partition,
each endpoint of an edge $e \in C_i \cap E_{i,j}$ 
must be in different color classes
in two bicliques that cover $e$.
For this purpose, we pay attention to the ordering of the roots of the stars
in making the 
partitions of $\widetilde{G}_{C_i}$ and $\widetilde{G}_{E_{i,j}}$.

For $\widetilde{G}_{C_i}$, we pick the root $u$ of the 
stars in the lexicographic order on the $n$-ary string
$(u_{k+i}u_{k+i-1} \cdots u_{i+1})$;
whereas for $\widetilde{G}_{E_{i,j}}$, we pick them in the reverse 
on the $n$-ary string \linebreak
$(u_{k+i}u_{k+i-1} \cdots u_{k+j} u_j u_{j-1} \cdots u_1)$.
In fact, we should only ensure that the $(k+i)$-th coordinate
is the most significant.
This guarantees that,
for every edge $\{u, v\} \in C_i \cap E_{i,j}$ with $u_{k+i} < v_{k+i}$,
$u$ is in the first class of a biclique in the collection for $G_{C_i}$
and in the second class of a biclique in the collection for $G_{E_{i,j}}$.
In this way, we have
\begin{eqnarray*}
	{\rm bp}_{1.5}(K_{n^{2k-1}})\leq (2k-1)\cdot n^{k-1}(n-1) + \sum_{i=1}^{k-1} i n^i(n-1) = O(kn^k).
\end{eqnarray*}
\Qed

By putting $k:=n$ in Theorem \ref{th:main}, the following is
immediate.

\begin{collorary}
\label{col:1}
$\bp_{1.5}(K_n) = n^{1/2+o(1)}$.
\end{collorary}

\proof
Let $N=n^{2k-1}$ and $k=n$.
A simple calculation shows
\begin{eqnarray*}
	n &=& \frac{\log N}{2\log n} + \frac{1}{2}\\
  	&=& \frac{\log N}{2\log (\frac{\log N}{2\log n}+\frac{1}{2})} + \frac{1}{2}\\
    &=& \frac{\log N}{2\log(\log N + \log n) - 2\log(2\log n)} + \frac{1}{2}\\
    &=& \Theta\left(\frac{\log N}{\log \log N}\right).
\end{eqnarray*}

By Theorem \ref{th:main}, we have
\begin{eqnarray*}
	\bp_{1.5}(K_N)=O(kn^k) &=& O(n^{n+1})
				= O(N^{\frac{n+1}{2n-1}})
        = O(N^{\frac{1}{2}+\frac{3}{4n-2}})\\
%  		&=& N^{\frac{1}{2}+\Theta\left(\frac{\log \log N}{2\log N - \log \log N}\right)}\\
  		&=& N^{\frac{1}{2}+\Theta\left(\frac{\log \log N}{\log N}\right)}\\
      &=& N^{\frac{1}{2}+o(1)}.
\end{eqnarray*}
\Qed

\medskip
An almost quadratic separation between
rank and fooling set size for $0/1-$matrices is 
immediately follows from Theorem \ref{th:main}. 

\begin{theorem}
\label{th:2}
There is a $0/1$ matrix $M$ such that
$\fool(M) \geq \rank(M)^{2-o(1)}$. 
\end{theorem}

\proof
\quad
Let $k:=n$ and $N:=n^k$.
Let $\{{\mathcal B}(U_i,W_i)\}_{i=1}^{m}$ be an
ordered biclique partition of $K_{N}$ constructed in Theorem \ref{th:main}. 
Let $A_i$ $(1 \leq i \leq m)$ be an $N \times N$ 
0/1-matrix whose $(k,\ell)$-entry is 1 iff $k \in U_i$ and $\ell \in W_i$ and
%let $M:= \sum_{i} A_i$, i.e., 
$M$ be the component-wise sum of all $A_i$'s.
Obviously, $M$ is a $0/1$-matrix of rank at most $m=N^{1/2+o(1)}$
since the rank of $A_i$ is 1 for all $i$.
In addition, 
the set of all the diagonal entries of $M$ forms a fooling set of $M$  
since all the diagonal entries of $M$ are zero and, 
for every $k \neq \ell \in [N]$,
at least one of $M_{k,\ell}$ or $M_{\ell,k}$ is one.
This completes the proof of the theorem.
\Qed

\bigskip
Indeed,
we constructed an $N \times N$ matrix having 
$0$-fooling set of size $N$ such that 
all one entries can be covered by
$N^{1/2+o(1)}$ disjoint 1-monochromatic rectangles.
As noted in \cite[Section 2.2]{Ama13}, this also yields a separation
between the {\it deterministic} and {\it unambiguous nondeterministic} 
communication complexities introduced by Yannakakis \cite{Yan91}. 
See \cite{Ama13} for more details.

By an equivalence between the problem to finding an ordered
biclique partition and the one to obtaining a lower bound
on the non-deterministic communication complexity
for the clique vs. independent set problem described in Introduction,
Theorem \ref{th:main} also implies the following:

\begin{theorem}
\label{th:3}
There exists an infinite family of graphs $G=(V(G), E(G))$ such that
the non-deterministic communication complexity of the clique vs.\ independent set problem is at least $(2-o(1))\log_2 |V(G)|$.
\Qed
\end{theorem}

\section{Concluding Remarks}
In this note, we established an almost tight bound on $\bp_{1.5}(K_n)$.
It is now known that
\begin{eqnarray*}
\Theta(n^{1/2}) = \bp_{2}(K_n) \leq \bp_{1.5}(K_n) = n^{1/2+o(1)}.
\end{eqnarray*}
It would be interesting to see whether $o(1)$ term in the exponent
can be removed or not.
A table of $\bp_{2}(K_n)$ and $\bp_{1.5}(K_n)$ for small values of 
$n$ $(n \leq 11)$ was shown in \cite[Section 3]{Ama13}.

More challenging problem is to find a graph
that has a larger (than quadratic) gap 
between its chromatic number and ordered biclique partition size.
A superpolynomial gap on them gives $\omega(\log |V(G)|)$ lower
bounds on the nondeterministic communication complexity of the
clique vs.\ independent set problem,
which would resolve a long standing open problem.

\end{document}